\documentclass[cameraready]{Interspeech}
\usepackage{multirow}
\usepackage[dvipsnames]{xcolor}

\title{QAMO: Quality-aware Multi-centroid One-class Learning\\ For Speech Deepfake Detection}

\author[affiliation={1}, orcid=0009-0002-1767-7598]{Duc-Tuan}{Truong}
\author[affiliation={2}, orcid=0000-0003-3472-0703, correspondingauthor]{Tianchi}{Liu}
\author[affiliation={2}, orcid=0000-0003-0021-5661, correspondingauthor]{Ruijie}{Tao}
\author[affiliation={3}, orcid=0009-0007-9654-9223, ]{Junjie}{Li}
\author[affiliation={3}, orcid=0000-0001-9133-3000, ]{Kong Aik}{Lee}
\author[affiliation={1}, orcid=0000-0001-6257-7399]{Eng Siong}{Chng}

\address{
    $^1$ Nanyang Technological University, Singapore \quad
    $^2$ National University of Singapore, Singapore \\
    $^3$ The Hong Kong Polytechnic University, Hong Kong
}

\email{truongdu001@e.ntu.edu.sg, tianchi\_liu@u.nus.edu, ruijie.tao@u.nus.edu, junjie98.li@connect.polyu.hk, kong-aik.lee@polyu.edu.hk, aseschng@ntu.edu.sg}


\usepackage{comment}

\begin{document}

\maketitle

\begin{abstract}
Recent work shows that one-class learning (OCL) can detect unseen deepfake by modeling a compact distribution of bona fide speech around a single centroid. However, OCL can oversimplify the bona fide class distribution and ignore useful cues, such as speech quality, which reflects the naturalness of the speech. Speech quality can be easily estimated via the Mean Opinion Score using assessment models. In this paper, we propose QAMO: Quality-Aware Multi-Centroid One-Class Learning for speech deepfake detection. QAMO improves OCL by introducing multiple quality-aware centroids. Each centroid is optimized to represent distinct speech quality subspaces to better capture the intra-class variability of bona fide speech. In addition, QAMO supports a multi-centroid ensemble scoring strategy that improves decision threshold and removes the need for quality labels at inference. QAMO achieves a 5.21\% equal error rate in the In-the-Wild dataset, outperforming prior OCL and quality-aware systems\footnote{Code is available at \href{https://github.com/ductuantruong/QAMO}{github.com/ductuantruong/QAMO}}.
\end{abstract}

\section{Introduction}
\label{sec:intro}
Traditional approaches treat speech deepfake detection (SDD) as a binary classification task, where models are trained to distinguish between real and fake speech \cite{kan,kwok2025robust,mask_aug,wedefense,mingli, gada,fg_add}. However, these methods tend to overfit known spoof attacks and reduce their ability to detect unseen ones \cite{cross_lingual,robust_media,noise_robust_kr,oc_acs,samo,other_oc}. To overcome this, one-class learning \cite{ocsoftmax} has been proposed as an alternative training scheme. Instead of learning distinct representations for bona fide and spoofed speech, one-class learning focuses on modeling a compact space of real speech around a single centroid and considers deviations from this centroid as potential spoofed speech. By focusing on the genuine speech, this approach generalizes better to unknown deepfake attacks.

Despite its advantages, conventional one-class learning constrains bona fide speech to a single compact unimodal distribution, which can oversimplify the diverse nature of genuine speech. Evidence from SAMO \cite{samo} shows that multi-centroid modeling effectively represents intra-class speaker-related characteristics, leading to improved performance over single-centroid one-class learning baselines. Beyond speaker information, another important speech factor is speech quality that can influence the separation between bona fide and spoofed speech. Speech quality reflects the perceived naturalness and overall listening experience of an utterance (e.g. clarity, absence of distortions/noise, and the realism of prosody). The Mean Opinion Score (MOS) is a widely used subjective quality metric that summarizes human-rated speech quality on an ordinal scale and is often approximated using learned predictors \cite{mos_survey}. The study reported in \cite{diff_mos_dataset} suggests that there are MOS gaps between real and synthetic speech in existing speech deepfake detection datasets. In particular, it was shown in \cite{nacl,other_quality} that the use of MOS in the selection of training samples improves the performance of SDD.


Motivated by these findings, we propose QAMO, a quality-aware multi-centroid one-class learning framework for speech deepfake detection. Unlike the conventional one-class learning assumes that a single centroid can represent the bona fide subspace, QAMO introduces multiple centroids, each associated with a distinct speech quality level. These centroids are optimized through a quality-level classification objective, where quality level labels are obtained by grouping the given MOS into discrete levels. Hence, QAMO explicitly encodes quality information into the bona fide representation, preserving intra-class variability while maintaining discriminability. Furthermore, QAMO does not require quality labels during inference since it computes the prediction score by ensembling distances across all quality-aware centroids. Extensive experiments show that QAMO significantly improves performance over baselines in various SDD benchmarks.

\section{Methodology}
\label{sec:format}
\begin{figure}[t]
  \centering
  \includegraphics[width=0.97\linewidth]{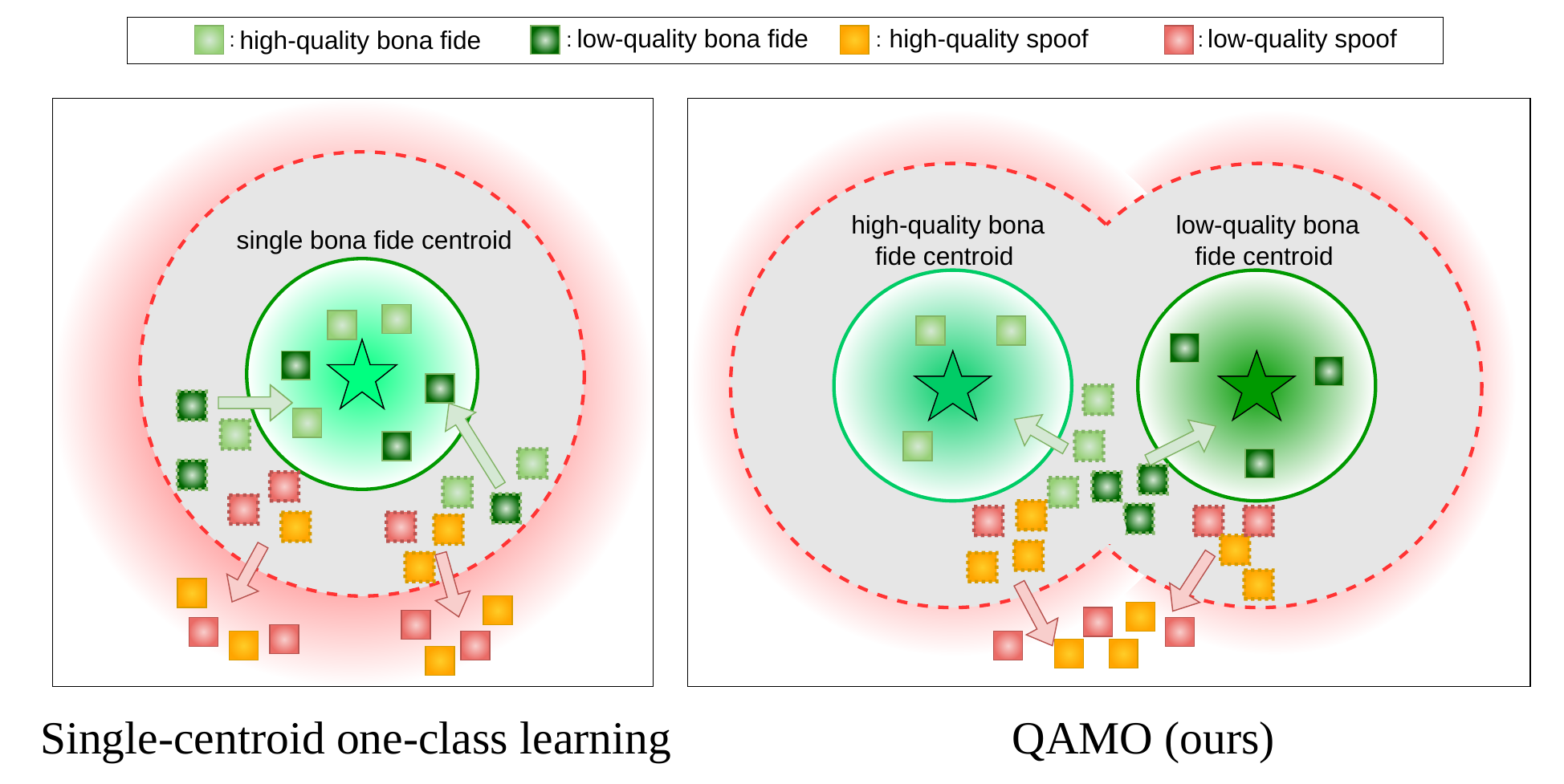}
 \vspace{-0.2cm}
  \caption{Illustration of the single-centroid one-class learning and our proposed QAMO. The gray region indicates the subspace between the bona fide and spoof margins.}
  \label{fig:main}
\end{figure}
This section introduces the proposed Quality-Aware Multi-Centroid One-Class Learning (QAMO) framework for speech deepfake detection. Figure \ref{fig:main} provides an overview of QAMO compared to the single-centroid one-class model, OC-Softmax \cite{ocsoftmax}. In OC-Softmax, bona fide embeddings are pulled toward a single centroid, while spoof embeddings are separated by margin boundaries. In contrast, QAMO uses multiple centroids, each corresponding to a distinct bona fide quality level, to capture intra-class variation in the bona fide speech representation.

Unlike the earlier multi-centroid one-class model SAMO \cite{samo}, which organizes embeddings by speaker identity, QAMO structures them by speech quality levels. Quality labels can be easily obtained from a speech assessment model, whereas speaker identity is often unavailable. QAMO learns discriminative quality-aware centroids through a classification objective. In contrast, SAMO constructs centroids by averaging embeddings from the same speaker, which risks the centroids to collapse into a single point since they lack explicit classification supervision.

\subsection{Quality-Aware Multi-Centroid modeling}

To capture the quality level of training samples, QAMO first assigns a discrete quality level to each sample using MOS predicted by a speech quality model \cite{scoreq}. The MOS values are then thresholded to form $Q$ discrete quality classes. In this study, we partition the MOS range into $Q=\{0, 1\}$ levels, denoting low (low MOS) and high (high MOS) quality, respectively, as:
\begin{equation}
    q_i = 
    \begin{cases}
    0, & \text{if}\ \mathrm{MOS}(x_i) < \tau \\
    1, & \text{if}\ \mathrm{MOS}(x_i) \geq \tau 
    \end{cases}
    \end{equation}
where $\tau$ is a predetermined threshold and $q_i$ denotes the quality level label for the utterance $x_i$.

In QAMO, each bona fide centroid of the multi-centroid is a learnable embedding $\mathbf{w}_q \in \mathbb{R}^D$ associated with a quality level $q \in Q$. These centroids are trained exclusively on bona fide embeddings with the corresponding quality level. To ensure that the multi-centroid embeddings are discriminative and aligned with its corresponding quality level, we employ the AM-Softmax loss \cite{amsoftmax} for quality classification:
\begin{equation}
\mathcal{L}_{\text{quality}} =-\frac{1}{\mathcal{B}} \sum_{i=1}^\mathcal{B} \log \frac{e^{s\left(\mathbf{w}_{q_i}^\top \hat{\mathbf{x}}_i-m\right)}}{e^{s\left(\mathbf{w}_{q_i}^T \hat{\mathbf{x}}_i-m\right)}+\sum_{j \neq q_i} e^{s\left(\mathbf{w}_j^T \hat{\mathbf{x}}_i\right)}}
\label{eq:cls}
\end{equation}
where $\mathcal{B}$ is the number of bona fide samples, $q_i$ is the quality level for the $i$-th sample, $\hat{\mathbf{x}}_i$ denotes the normalized embedding of the utterance $i$, $m$ is the additive margin, and $s$ is a scale factor.

\subsection{Quality-Aware Multi-Centroid One-Class learning}
QAMO extends the single-centroid OC-Softmax loss \cite{ocsoftmax}  to multiple centroids that account for different levels of speech quality. For a bona fide input, the cosine similarity distance is computed with respect to the centroid $q$ of their quality level. For a spoof input, the maximum cosine similarity across all centroids is penalized to ensure separation from the bona fide. Let $y_i$ denote the detection class ($1$ for spoof, $0$ for bona fide), the similarity distance is defined as:
\begin{equation}
d_i =
\begin{cases}
\mathbf{w}_{q_i}^{\top} \hat{\mathbf{x}}_i, & \text{if}\ y_i = 0 \\
\max\limits_q\left( \mathbf{w}_{q}^{\top} \hat{\mathbf{x}}_i \right), & \text{if}\ y_i = 1
\end{cases}
\end{equation}
The QAMO loss is then formulated as:
\begin{equation}
    \mathcal{L}_{\text{QAMO}} = \frac{1}{N} \sum_{i=1}^{N}
\log(
1 + e^{ \alpha (m_{y_i} - d_i)(-1)^{y_i} })
\label{eq:qamo}
\end{equation}
where $\alpha$ is a scaling factor, $m_{0}$ and $m_{1}$ are the margins for bona fide and spoof classes, and $N$ is the mini-batch size. The final training objective combines the proposed QAMO loss (Eq.~\ref{eq:qamo}) with the quality classification  (Eq.~\ref{eq:cls}) :
\begin{equation}
    \mathcal{L} = \mathcal{L}_{\text{QAMO}} + \lambda \mathcal{L}_{\text{quality}}
\end{equation}
where $\lambda$ is a weight hyperparameter. 

\subsection{Inference with QAMO}
At inference stage, QAMO produces a countermeasure (CM) score by computing the similarity distance $d_i$ between input utterances $x_i$ and the learned centroids $\mathbf{w}_{q}$. If the quality label $q$ is available, the score is simply the cosine similarity to its corresponding centroid, providing a direct quality-aware decision. However, this requires an automatic MOS predictor to estimate quality, which increases the computational cost in real-life deployment. To avoid this, QAMO can operate without quality labels, similar to SAMO \cite{samo}, by taking the maximum similarity across all centroids (max-score inference):
\begin{equation}
    d_i=\max _q\left( \mathbf{w}_{q}^{\top} \hat{\mathbf{x}}_i \right)
\end{equation}
This design is effective because the centroids are optimized with a quality classification loss which aligns each utterance embedding to their corresponding quality centroid. However, a max-score strategy is equivalent to a hard assignment to a single centroid, making it sensitive to quality estimation errors under unseen degradations or ambiguous conditions. In addition, spoofed samples may spuriously align with one centroid, resulting in overconfident predictions. To achieve more stable and uncertainty-aware inference, QAMO adopts a softmax-weighted ensemble scoring strategy (ensemble-score inference). The final CM score is computed as a weighted sum of similarities to all centroids:
\begin{equation}
    d_i= \sum_{q\in Q}\alpha_q\left(\mathbf{w}_{q}^{\top} \hat{\mathbf{x}}_i\right)
\end{equation}
where weights $\alpha_q$ are obtained by applying a softmax to cosine similarities:
\begin{equation}
    \alpha_q= \frac{e^{\left(\mathbf{w}_{q}^\top \hat{\mathbf{x}}_i\right)}}{\sum_{j \in Q}e^{\left(\mathbf{w}_{j}^T \hat{\mathbf{x}}_i\right)}}
\end{equation}
Although the centroids correspond to different quality levels, they jointly represent the bona fide class. This ensemble formulation emphasizes the most compatible quality level while aggregating evidence across all centroids, leading to smoother, lower-variance, and better-calibrated scores under quality uncertainty. The additional computational cost is negligible, as it reuses the precomputed centroid similarities and only introduces a softmax operation and weighted summation.


\begin{table*}[t!]
\centering
\caption{Performance of QAMO compared with baseline and recent methods across multiple test sets. Results marked with * are reported from \cite{arena}. Values in parentheses denote results reproduced under our experimental settings, which differ due to the change of data augmentation configuration described in section \ref{subsec:setting}.}
\begin{tabular}{lccccc}
\hline
\multicolumn{1}{c}{\multirow{2}{*}{\textbf{System}}} & \multirow{2}{*}{\textbf{\begin{tabular}[c]{@{}c@{}}Loss\\ Function\end{tabular}}} & \multicolumn{4}{c}{\textbf{EER$\downarrow$ (\%)}}                                \\ \cline{3-6} 
\multicolumn{1}{c}{}                                 &                                                                                   & \textbf{21LA}   & \textbf{21DF}   & \textbf{ITW}    & \textbf{FoR}  \\ \hline
XLSR-Conformer \cite{conformer}                                               & \multirow{4}{*}{WCE}                                                              & 0.97 & 2.58 & 8.42 & -                            \\
ASC+OC \cite{oc_acs}                                             &                                                                                   & 1.30   & 2.19   & -   & -                              \\
XLSR-Mamba \cite{xlsr_mamba}                                       &                                                                                   & 0.93   & 1.88   & 6.70   & 6.71*                               \\
XLSR-Conformer-NACL \cite{nacl}              &                                                                                   & 0.89   & 1.88   & 6.60   & -                                 \\  \hline

\multirow{3}{*}{XLSR-Nes2NetX \cite{nes2net}}                       & WCE                                                                               & 2.00 (3.9)   & 1.78 (2.76)   & 6.60 (9.76)   & 6.31* (10.12)                              \\
                                                     & OC-Softmax                                                                                 & 3.36 & 2.29 & \textbf{8.23} & \textbf{3.97}                            \\
                                                     & QAMO (ours)                                                                              & \textbf{2.29} & \textbf{1.60}  & 8.83 & 4.90                            \\ \hline
\multirow{3}{*}{XLSR-Conformer-TCM \cite{conformer_tcm}}                  & WCE                                                                               & \textbf{1.03 (1.37)}   & 2.06 (2.39)   & 7.79 (7.13)   & 10.68* (5.70)                             \\
                                                     & OC-Softmax                                                                                & 1.75 & 1.89 & 6.72 & 5.65                            \\
                                                     & QAMO (ours)                                                                              & 2.53 & \textbf{1.63} & \textbf{5.21} & \textbf{3.45}                            \\ \hline
\end{tabular}
\label{tab:diff_arch}
\end{table*}

\section{Experiments and results}
\label{sec:experiment}
\subsection{Dataset and metrics}
\begin{figure}[t]
\centering
\includegraphics[width=0.9\linewidth]{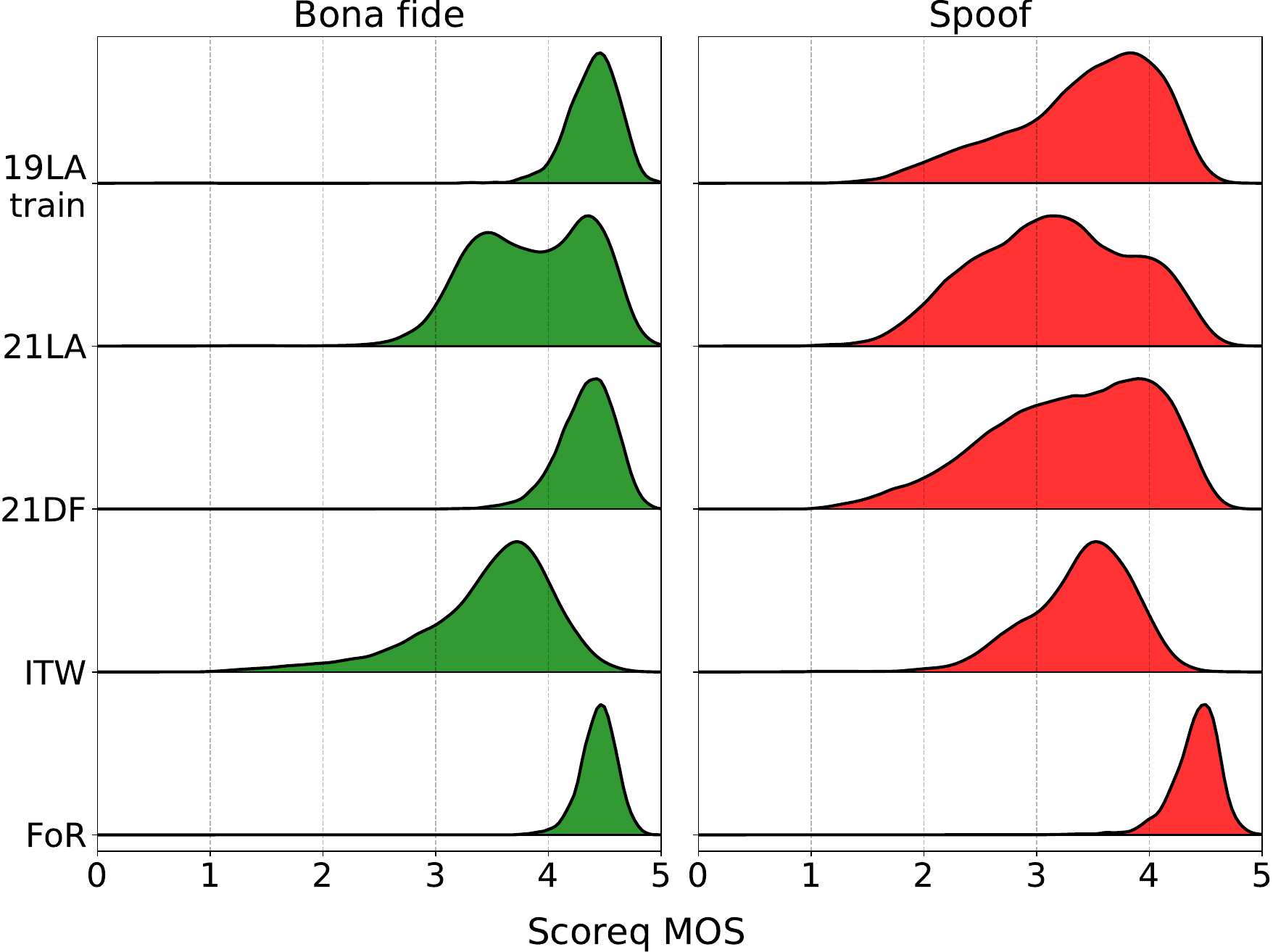}
\vspace{-0.2cm}
\caption{MOS distributions across our examined datasets.}
\label{fig:score_data}
\end{figure}
We train and validate our models on the ASVspoof2019 Logical Access (LA) dataset \cite{asvspoof2021}. To evaluate its robustness to unseen deepfake attacks and a variety of acoustic conditions, we test it on four datasets: ASVspoof2021 LA, ASVspoof2021 DF \cite{asvspoof2021}, In-the-Wild (ITW) \cite{itw}, and the Fake-or-Real (FoR) \textit{norm-test} subset\footnote{\href{https://www.kaggle.com/datasets/mohammedabdeldayem/the-fake-or-real-dataset}{kaggle.com/datasets/the-fake-or-real-dataset}}. As shown in Figure \ref{fig:score_data}, these test sets also cover a wide range of speech quality levels for both bona fide and spoofed speech. We use the Equal Error Rate (EER) as the main metric for performance evaluation.
\subsection{Implementation details}
\label{subsec:setting}
Speech quality is estimated by the recent Scoreq MOS predictor \cite{scoreq}\footnote{\href{https://github.com/alessandroragano/scoreq}{github.com/alessandroragano/scoreq}}, with a threshold $\tau = 2.5$ to separate high and low quality levels. QAMO hyperparameters are set as $\alpha=20$, $m_0=0.9$ for bona fide and $m_1=0.2$ for spoof. For $\mathcal{L}_{\text{quality}}$, we use a scale value $s=20$, margin $m=0.4$ and its weight $\lambda=0.1$.  We apply QAMO on XLSR-TCM-Conformer \cite{conformer_tcm} and XLSR-Nes2NetX \cite{nes2net}, following the original training settings of the former. While previous works train separate models \cite{nacl, nes2net,conformer_tcm,pan24c_interspeech} with RawBoost \cite{rawboost} configurations 3 and 5 for ASVspoof2021 LA and DF evaluations, respectively, we unify our experiments by using RawBoost configuration 4 for data augmentation. This setup reduces the number of experiments while covering all noise types found in both configurations. 

\begin{figure}[t]
\centering
\includegraphics[width=0.85\linewidth]{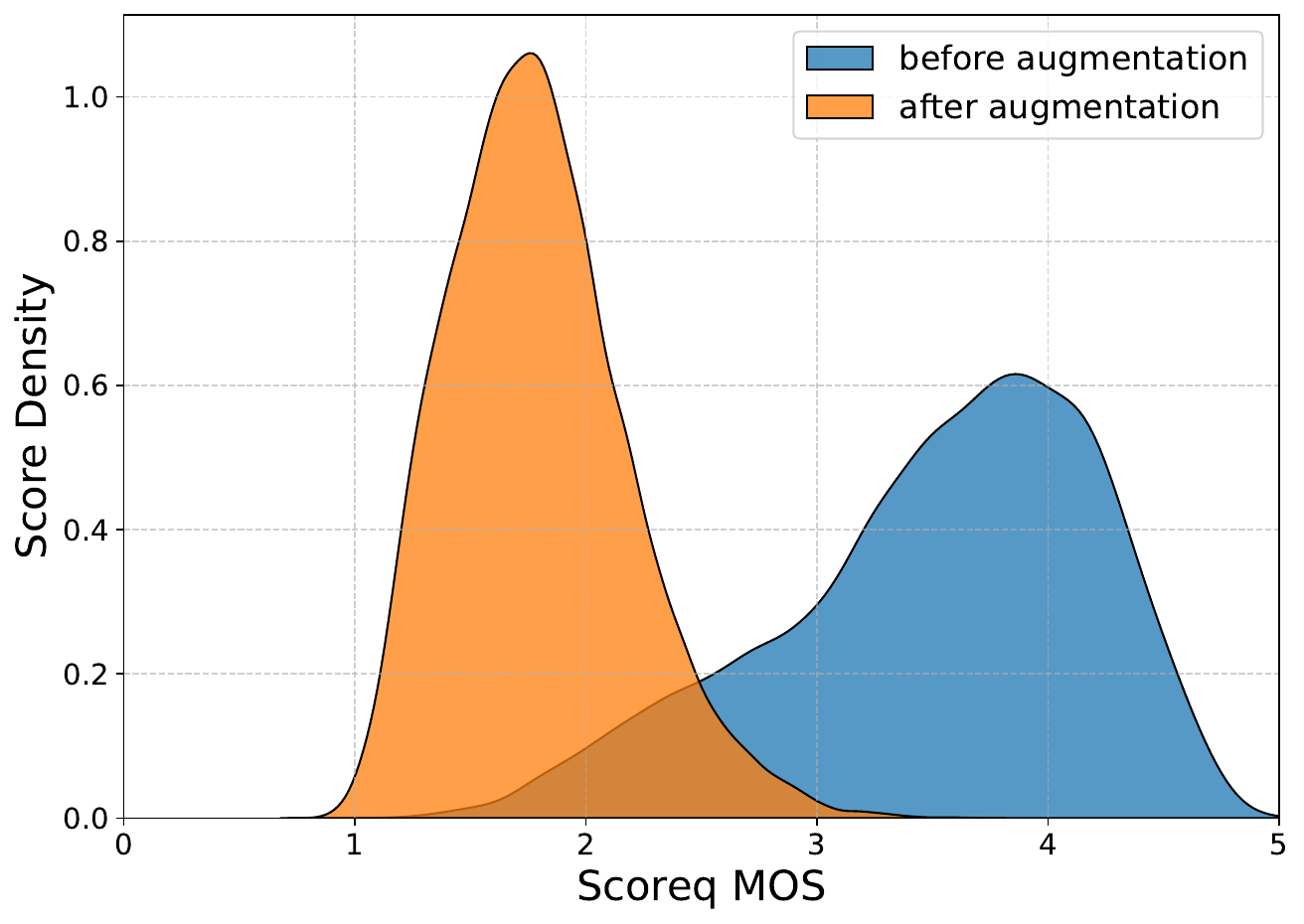}
\caption{MOS distributions of ASVspoof2019LA training samples before and after applying Rawboost augmentation.}
\label{fig:mos_aug}
\end{figure}

Unlike previous work \cite{nacl} that ignores quality degradation under data augmentation, we treat augmented inputs as low quality to reflect the expected MOS drop caused by additive noise and channel distortions. Our data augmentation is applied randomly on-the-fly at each training iteration, so the same utterance can be transformed differently across epochs. Computing an MOS predictor for every augmented waveform would introduce substantial overhead and significantly slow down the training. Therefore, we use a lightweight approximation and assign augmented samples to the low-quality group during training. Importantly, this is not an arbitrary assumption, as Fig.\ref{fig:mos_aug} shows that RawBoost augmentation causes a clear shift to the left in the predicted MOS in ASVspoof2019 LA training utterances, with the majority of augmented samples falling below the threshold ($\tau = 2.5$). Hence, the low-quality assignment is a compute-efficient proxy that is consistent with the MOS distribution induced by our augmentation pipeline. Since ASVspoof2019 LA contains mostly high-quality samples, we augment 40\% of training data to balance quality levels. No augmentation is applied during validation.

\begin{figure*}[t]
    \centering
    \includegraphics[width=0.99\linewidth]{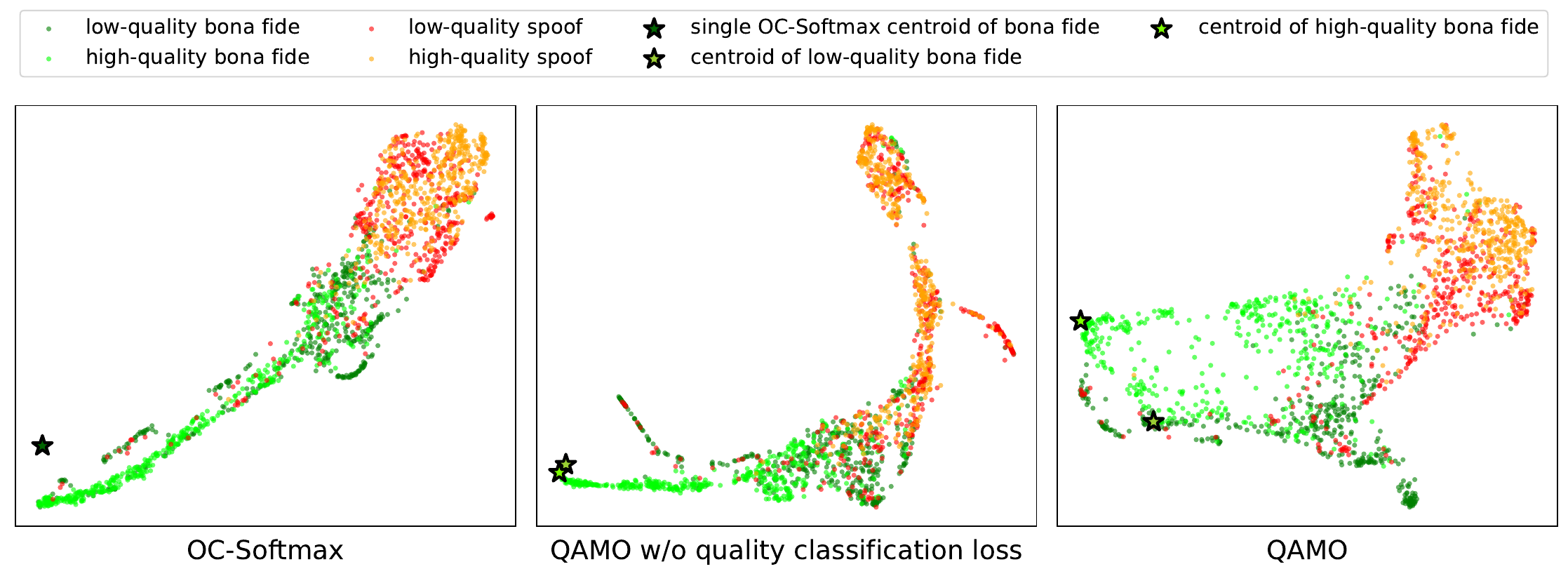}
    \vspace{-0.2cm}
    \caption{2D UMAP \cite{umap} feature embedding visualization of ITW samples from different trained XSLR-Conformer-TCM models.}
    \label{fig:umap_viz}
\end{figure*}
\section{RESULTS AND ANALYSIS}
\label{sec:result}
\subsection{Results of the proposed method}
Table \ref{tab:diff_arch} presents the performance of QAMO compared to baseline and recent models on four benchmarks. When integrated with XLSR-Nes2NetX, QAMO significantly reduces the EER from 3.36\% of OC-Softmax to 2.29\% in 21LA and from 2.29\% to 1.60\% in 21DF, although OC-Softmax remains slightly better in ITW and FoR. The improvements are clearer with XLSR-Conformer-TCM, where QAMO achieves the best results on 21DF (1.63\%), ITW (5.21\%), and FoR (3.45\%), outperforming both the weighted cross-entropy (WCE) loss, the single-centroid OC-Softmax, and the prior quality-aware model XLSR-Conformer-NACL. These results demonstrate that explicitly modeling multiple speech quality levels within one-class learning enhances robustness and yields more balanced detection performance across diverse evaluation conditions.
\subsection{Ablation study}
Table \ref{tab:abl_study} summarizes the results of different components in QAMO. First, adding the quality classification loss $\mathcal{L}_{\text{quality}}$ to the conventional weighted cross-entropy (WCE) yields only marginal gains and does not match the performance of QAMO, highlighting that simple quality conditioning is insufficient. Within QAMO, removing $\mathcal{L}_{\text{quality}}$ leads to performance degradation in 21DF and ITW, with results similar to OC-Softmax in Table \ref{tab:diff_arch}. This likely occurs because, without explicit quality classification supervision, the multiple centroids may collapse toward a single point. Finally, using maximum-score inference produces poorer results compared to ensemble-score inference, confirming that softmax-weighted sum across centroids provides a better scoring strategy.

\begin{table}[t]
\centering
\caption{Ablation study on different components of QAMO.}
\label{tab:abl_study}
\vspace{-1mm}
\begin{tabular}{lccc}
\hline
                                                        & \multicolumn{3}{c}{\textbf{EER (\%)}}         \\ \cline{2-4} 
                                                        & \textbf{21DF}          & \textbf{ITW}           & \textbf{FoR}           \\ \hline                          
WCE               & 2.39        &  7.13       & 5.70       \\
\quad + $\mathcal{L}_{\text{quality}}$    & 1.72        & 7.18       & 7.55       \\ \hline
QAMO              & \textbf{1.63}        & \textbf{5.21}       & 3.45       \\ 
\quad w/o $\mathcal{L}_{\text{quality}}$      & 2.17        & 6.47       & \textbf{2.78}       \\
\quad w/ max-score inference      & 2.29        & 6.31       & 5.16       \\ \hline
\end{tabular}
\end{table}

\subsection{Visualization of embedding and score distribution}
\label{subsec:viz}
To analyze the proposed method, we visualize the feature embeddings of samples in ITW test set, which serves as an out-of-domain evaluation. Figure \ref{fig:umap_viz} compares three training settings of the XLSR-Conformer-TCM backbone: (i) OC-Softmax, (ii) QAMO without $\mathcal{L}_{\text{quality}}$, and (iii) the full QAMO. In all settings, bona fide and spoofed samples form distinct regions, indicating that one-class learning remains effective under a domain shift. However, OC-Softmax and QAMO without $\mathcal{L}_{\text{quality}}$, show strong overlap among bona fide samples of different quality levels, and in the latter, the quality centroids collapse into a single point as hypothesized. In contrast, full QAMO separates bona fide samples by quality and aligns them with their respective centroids, preserving intra-class quality variation while maintaining a coherent bona fide subspace.

\begin{figure}[t]
  \centering
  \includegraphics[width=\linewidth]{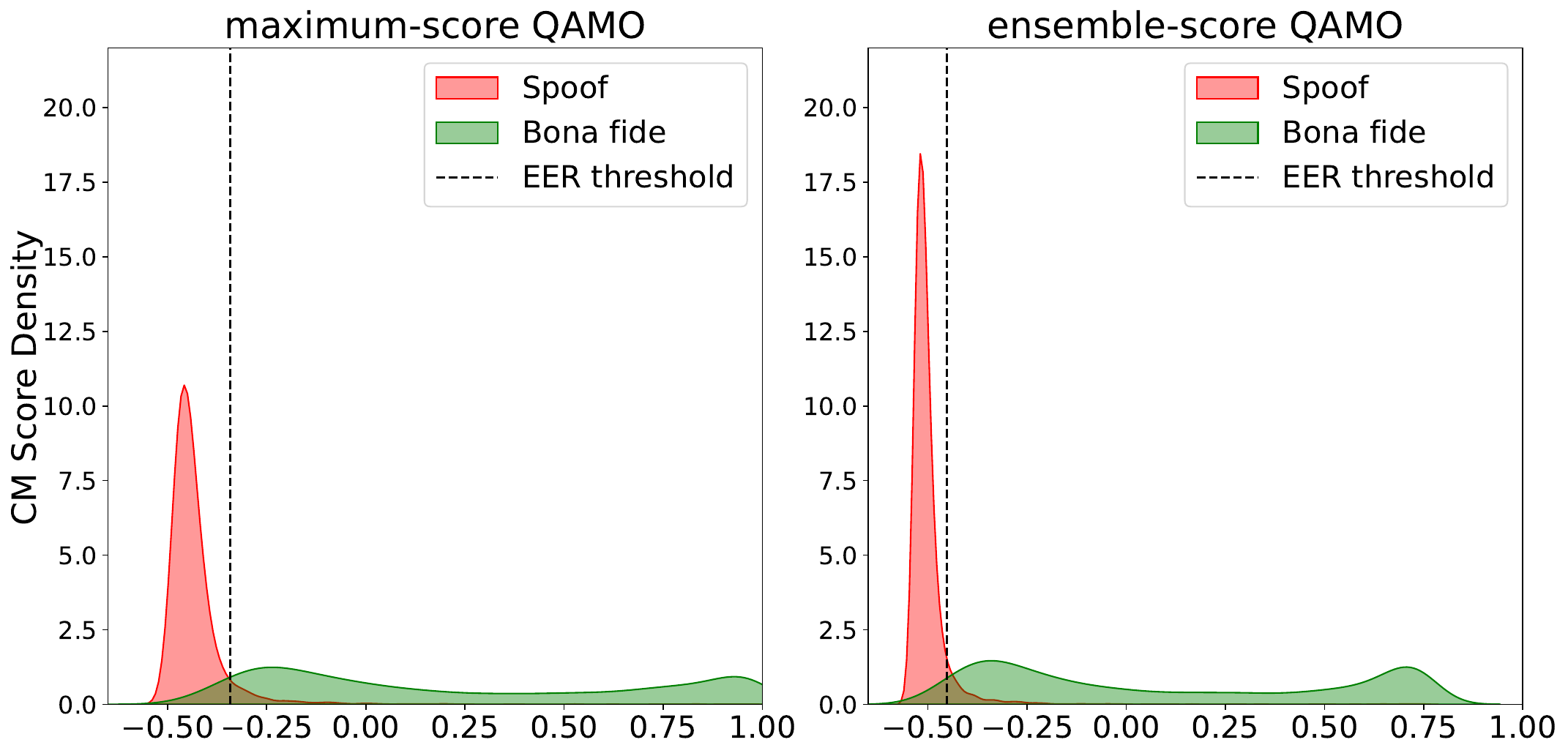}
  \vspace{-0.2cm}
  \caption{ITW score distributions of XLSR-Conformer-TCM with max-score and ensemble-score inference strategies.}
  \label{fig:score_dist}
\end{figure}

We visualize score distributions on the ITW test set under two inference strategies: maximum-score and ensemble-score. As shown in Figure \ref{fig:score_dist}, the maximum-score yields flatter, more overlapping distributions, making it difficult to obtain a clear threshold. In contrast, ensemble-score produces sharper separation between bona fide and spoof classes, offering a more stable and discriminative scoring space. This result justifies the use of ensemble scoring in QAMO, which normalizes predictions across centroids and enables more reliable threshold selection.

\section{Conclusion}
\label{sec:conclusion}
We have introduced QAMO, a quality-aware multi-centroid one-class learning for speech deepfake detection. By modeling bona fide speech with multiple quality-aware centroids, QAMO effectively preserves intra-class variability while enhancing discrimination against spoofed audio. Experiments on multiple benchmarks demonstrated that QAMO outperforms the conventional one-class baseline and other quality-aware methods, achieving a strong generalization to unseen attacks. Furthermore, the ensemble-score inference strategy was shown to stabilize decision boundaries and improve detection robustness. These findings highlight the importance of incorporating speech quality for building more reliable countermeasures against new deepfake attacks.

\newpage
\section{Generative AI Use Disclosure}
Generative AI tools were used to assist
with grammar correction and polishing of the manuscript. All
scientific content, experimental design, and analysis were conducted solely by the authors. All authors take full responsibility
for the content of this paper.
\bibliographystyle{IEEEtran}
\bibliography{mybib}

\end{document}